\documentclass
{aa}

\usepackage{natbib}
\usepackage{graphicx}

\begin{document}

\title{On the mass of moderately rotating strange stars in the MIT
bag model and LMXBs}

\titlerunning{On the mass of bag-model stars}

\author{ 
  J.L. Zdunik$^1$,
  T. Bulik$^1$,
  W. Klu\'zniak$^1$,
  P. Haensel$^1$,
  D. Gondek-Rosi\'nska$^2$
}

\authorrunning{ Zdunik, et. al.
  }

\institute{$^1$Nicolaus Copernicus Astronomical Center, 
Bartycka 18, 00-716 Warszawa, Poland\\
$^2$D\'epartement d'Astrophysique Relativiste et de Cosmologie---UPR 176
du CNRS, Observatoire de Paris, F-92195 Meudon Cedex, France
}

\date{Received , Accepted }

\thesaurus{02.04.01, 08.14.1}

\maketitle

\begin{abstract}

We compute the maximum mass of moderately rotating strange stars
as a function of the strange quark mass, 
of the QCD coupling constant, $\alpha_c$,
 and of the bag constant (vacuum energy density),
$B$, in the MIT bag model of quark matter with lowest order quark-gluon
interactions. For a fixed value of $B$, the maximum stellar mass
depends only weakly on  $\alpha_c$, and is independent of this
coupling in the limit of massless  quarks. However, if it is the value of
the chemical potential of quark matter at zero pressure which is held
constant, for example at the value corresponding to the stability
limit of nucleons against conversion to quark matter, the  maximum
mass of the strange star is higher by up to 25\% for
$\alpha_c=0.6$, than for non-interacting quarks, and this may be relevant
in the discussion of kHz QPO sources.
The maximum mass of a non-rotating strange star could be sufficiently
high to allow an orbital frequency 
as low as  1.0 kHz in the marginally stable orbit.
However, for all $\alpha_c<0.6$, the stellar mass cannot exceed
$2.6M_\odot$ at any rotational period $\ge1.6\,$ms.
\end{abstract}

\keywords{dense matter -- equation of state -- stars: binaries: general
-- X-rays: stars}

\section{Kilohertz QPOs and the mass of LMXBs}

The discovery of kHz QPOs \citep{1996IAUC.6320....1S,1996IAUC.6319....1V} 
in several low-mass X-ray binaries (LMXBs)
has renewed interest in the maximum mass of neutron stars,
as its value limits the maximum observable orbital frequency
\citep{1990ApJ...358..538K}.
The maximum
masses for neutron stars modeled with various equations of state
(e.o.s.) are well established
\citep{1977ApJS...33..415A,1986ApJ...304..115F,1994ApJ...424..823C,Salgado},
and have been examined in the context of kHz QPOs \citep{1998ApJ...509L..37K}.
In principle, LMXBs could contain strange stars instead
(Cheng and Dai 1996\nocite{1996ApJ...468..819C}),
and it has been asked
whether the observed values of kHz QPOs
are compatible with the masses of such stars (Bulik et al., 1999a,b).
\nocite{1999A&A...344L..71B}

The existence of strange stars would
shed light on the issue of stability of quark matter.
It has been suggested that a quark fluid, composed of roughly equal number of
up, down and strange quarks, may be the ground state of bulk matter
\citep{1971Bodmer,1984Witten} and a detailed descripion of such 
strange matter has been given by \cite{FarJaf}. This idea has not been
universally accepted, and it was argued that for realistic values
of the QCD coupling constant, the phase transition to quark matter
would occur at densities too high to be of interest 
(e.g., Bethe, Brown and Cooperstein, 1987).
\nocite{Bethe}

Strange stars may be very hard to distinguish from
neutron stars, particularly in LMXBs, as they may have a crust of normal
matter \citep{1986ApJ...310..261A,1986A&A...160..121H}.
 The crust will contribute little to the mass ($\le 10^{-5}M_\odot$), but
is expected to have a thickness sufficient to support
nuclear burning, including helium flashes responsible for X-ray bursts,
thus mimicking neutron stars.
The heat released in conversion (at the bottom of the crust)
of nuclei into strange matter is directed into the
strange matter core
\citep{1990ApJ...362..572M,1990A&A...229..117H}.
At any rate, steady release of energy from nuclear conversion is
very difficult to distinguish from the gravitational binding energy released
in steady accretion.

However, in young radio pulsars the crust thickness does make 
a difference---the crust in strange stars would be far too thin to allow
the redistribution of angular momentum necessary to explain the
glitch phenomenon (sudden spin-up of a pulsar), therefore glitching
pulsars are though to be neutron stars and not quark stars \citep{Alpar}.
It is likely that the coalescence of two strange stars would
lead to the contamination of our Galaxy with chunks of strange matter,
and would thus preclude the formation of young neutron stars
\citep{Madsen,Caldwell}. These arguments make unlikely the presence of strange
stars in the population of ordinary pulsars (including binary Hulse-Taylor type
pulsars) or their accreting counterparts, the high-mass X-ray binaries.
However, the presence of strange stars among the millisecond pulsars
or LMXBs seems to be allowed \citep{Kluz94,1996ApJ...468..819C}.
In these old systems,
the stellar mass itself may yield clues as to
the nature of the compact object. 

We are led to consider the maximum mass of strange
stars in the expectation that LMXB masses will become available in
the near future, either through a better understanding of accretion
phenomena, including the observed quasi-periodic oscillations (QPOs) in
the X-ray light curve, or through classical optical determinations
of binary parameters for newly discovered transient sources.

The maximum frequency of stable orbital motion is attained in the
innermost (marginally) stable circular
orbit (ISCO) allowed by general relativity,
and if its value for some X-ray source
were that of the observed maximum QPO 
frequency---ranging from 1.0 kHz to 1.2 kHz in twelve LMXBs---conclusions
could be reached about the nature of the compact object.
[For a review of QPOs see e.g. \cite{QPOrev98}.]

The question, whether ISCO frequencies as low as the maximum observed
QPO frequencies can be attained
outside quark stars, has been answered in the affirmative for rapidly rotating
strange stars, with an equation of state (e.o.s.) based on the MIT bag model
of quark matter---strange stars rotating close to the equatorial mass-shedding
limit can have ISCO frequencies below 1~kHz for masses as low as
$1.4 M_\odot$ \citep{1999A&A...352L.116S}. 
However, the same question has not yet been answered for moderately rotating
strange stars, of periods $P\sim 3\,$ms or more, for which such ISCO
frequencies would imply larger masses: 
$M=2.2M_\odot(1+0.75j)(1.0\,{\rm kHz}/f_{max})$, where $f_{max}$ is the ISCO
frequency, and $j\le0.3$ is the dimensionless angular momentum
\citep{1990ApJ...358..538K,1998ApJ...509L..37K}.

There is a compelling reason
to consider stars rotating at relatively low rates.
The  period of the compact star
in the the transient LMXB SAX J~1808.4-3658 \citep{1998Natur.394..344W},
the first of possibly many such transients to be discovered,
has been measured to be 2.5 ms. Whether
or not kHz QPOs will be discovered in that source, its mass
may eventually be determined by optical studies of the binary companion.

It has also been argued that the oscillations seen in some X-ray bursters imply
a stellar rotational frequency of $\sim 300\,$Hz
\citep{1997ApJ...486..355S}. Finally, it is not yet
clear whether strange stars can attain the equatorial mass-shedding limit
because of the unusually high value of $T/W>0.2$ calculated for their
models---note that for Newtonian stars the secular instability
to a bar-mode deformation  sets in already at $0.1275\le T/W\le 0.1375$
\citep{1996ApJ...460..379B}. For a discussion of other modes,
also in general relativity,
see, e.g., Gourgoulhon et al. (1999), and references therein.

For all these reasons, we considered exact numerical models of strange
stars rotating with frequencies up to 700~Hz,
and found that, to within a few per cent,
the maximum mass of such strange stars is
the same as that of the static configurations.
Therefore, we investigate the maximum mass of static
(non-rotating) strange stars, which till now has
been discussed in the context of QPOs only in the simplest case of ideal
quark gas in the bag model  \citep{1999A&A...344L..71B}.
Here, we use the e.o.s. of interacting, massive quarks within the MIT bag
model of self-bound quark matter.

\begin{figure}
\includegraphics[angle=-90,width=0.95\columnwidth]{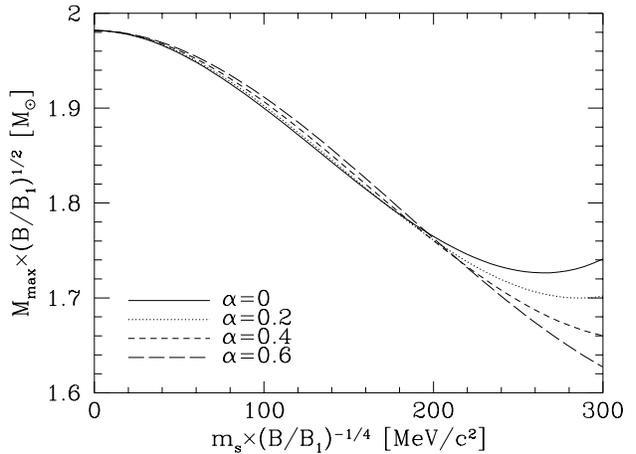}
\caption{The maximum mass of a strange star as a function of
the strange quark mass, for various values of the QCD coupling
constant. The stellar mass scales as the inverse square root
of the bag constant, $B$, provided that the quark mass is scaled
with $B^{1/4}$. Here,  $B_1=58.9\,$MeV$\,{\rm fm}^{-3}$.
}
\end{figure}

\section{Strange stars}

The maximum mass of quark stars was first derived \citep{1976Natur.259..377B}
for an unusually
low value of the bag constant.
\cite{1984Witten} showed that the maximum
mass of a static strange star is,
within the MIT bag model (Farhi and Jaffe 1984),
 
\begin{equation}
M_{max}=1.98M_\odot (B/B_1)^{-1/2},
\label{witt}
\end{equation}
where $B_1=58.9\,$MeV$\cdot$fm$^{-3}$.

There seems to have been no systematic investigation of the {\sl maximum} mass
of the star as a function of the three basic parameters of
quark matter in the MIT bag model:
the mass of the strange quark, $m_s$, the bag constant $B$, and the
strength of the QCD coupling constant, $\alpha_c=g^2/4\pi$.
Detailed models of  strange stars have been constructed
and the structure of strange stars, including the mass--radius relationship
has been discussed extensively in the literature
\citep{1986ApJ...310..261A,1986A&A...160..121H,1989PhRvL..63.2629G,1989Natur.341..633F,Prakash}, however discussion of stellar parameters tended to
concentrate on the maximum value of the bag constant, setting a {\it lower}
bound on the maximum mass and an upper bound on the rotational frequency.

In this section, we discuss only static stellar models, constructed in general
relativity by solving the TOV equations
\citep{1938PhysRev...55..374}. 
We neglect the crust, whose contribution to the
maximum mass is quite minor for stars composed mostly of self-bound
quark matter
\citep{1986A&A...160..121H}.
Following Farhi and Jaffe (1984), we take $m_s$ and $\alpha_c$ to
be renormalized at $q=313\,$ MeV, and as a model for quark matter
consider a ``bag'' of positive vacuum energy density, $B$, filled
with quarks (of two massless flavors and one massive) having
interactions through first order in $\alpha_c$.  The actual form of
the thermodynamic expressions we use can be found in 
\citet{1986A&A...160..121H}.
 As both the
pressure and the density scale with the bag constant, TOV equations
imply that the stellar mass and radius scale as $M\propto
B^{-1/2}$, and $R\propto B^{-1/2}$ \citep{1984Witten}, provided that the
masses of the quarks scale as $m\propto B^{1/4}$.

In Figure~1, we plot the rescaled value of the maximum stellar mass,
$M(B/B_1)^{1/2}$, as a function of the rescaled strange quark mass,
$m_s(B_1/B)^{1/4}$, for various values of the QCD coupling
constant.
Note that the highest value of $M_{max}$, the maximum mass of the strange star,
is independent of $\alpha_c$, if $B$ is fixed, and
 is obtained for $m_s=0$, i.e., for massless quarks.
But in fact, the value
of $B$ is not known, and as its lower bound does depend on the
value of  $\alpha_c$, the actual physical bounds on the maximum
mass of a strange star will depend, through $B$, on the coupling
constant. This is discussed in the next section.

\section{The minimum value of $B$ and an upper bound to the mass
of moderately rotating strange stars}

To determine the highest possible value of the maximum
mass of static  strange stars in the MIT bag model, it is
enough to consider the e.o.s. of an ulrarelativistic Fermi gas in a
volume with vacuum energy density $B> 0$. If the quarks
are massive, the actual maximum mass of the star will be somewhat
lower---as is evident from Figure~1, at fixed values of the
other parameters, the maximum mass of a strange star decreases with
increasing quark mass.\footnote{The dependence of $M$ on $m_s$ is
easily understood in the limit $\alpha_c=0$. The mass of a static
star modeled with ideal Fermi gas e.o.s. scales with fermion mass as
$M\propto m^{-2} $ and attains a finite limit at $m\rightarrow 0$
\citep{1938PhysRev...55..374}.} 

Currently, the actual value of $B$ cannot be reliably derived from
fits to hadronic masses of the quark-model of nucleons.  Instead, we
must rely on a different argument to find $B_{min}$.  We require
that neutrons do not combine to form plasma of deconfined up and
down quarks, or equivalently, that quark matter composed of up and
down quarks in 1:2 ratio is unstable to emission of neutrons
through the reaction $u+2d \rightarrow n$
(e.g., Haensel, 1987; Farhi, 1991\nocite{Haen1987,Farhi1991}),
i.e., that the
baryonic
chemical potential at zero pressure of such quark matter satisfies 
\begin{equation}
\mu^{u,d}_0 > 939.57\,{\rm MeV}. 
\label{bmin}
\end{equation}
This condition provides a lower bound on the bag constant $B$,
and consequently, by eq.~(\ref{witt}),
an upper bound on the mass of static strange stars.
We take the up and down quarks to be massless. Therefore,
eq. (\ref{bmin}) implies $B\ge B_1(1-2\alpha_c/\pi)$.
 For noninteracting and massless quarks
($\alpha=0$, $m_s=0$),
 this value of the bag constant corresponds to a minimum
density of strange matter (attained at zero pressure) of
$\rho_0(0)=4.20\times 10^{14}\,$g/cm$^3\equiv 4B_1/c^2$,
with a corresponding maximum mass of non-rotating strange stars
of $M_{max}(0)=1.98M_\odot$. 

\begin{figure}
\includegraphics[angle=-90,width=0.95\columnwidth]{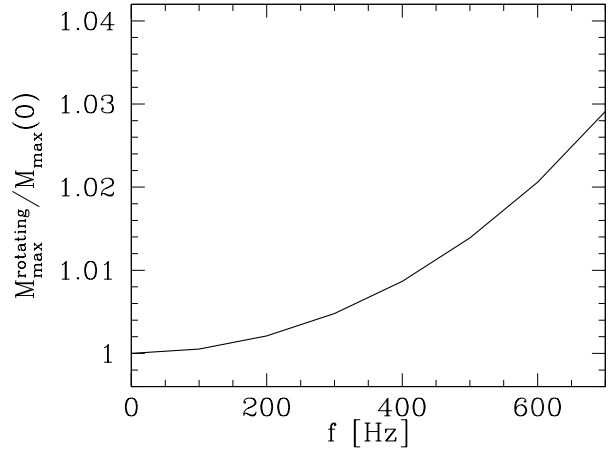}
\caption{The dependence on rotational frequency of the maximum mass of a
strange star, as derived from sequences of
fully relativistic numerical models for the e.o.s. of massless, noninteracting
quarks (in a bag). 
In the figure, the stellar mass is scaled with the
maximum mass of static models for this e.o.s., $1.98M_\odot$.
 Note, that for all periods of stellar
rotation in the currently observed range
($P\ge1.6\,$ms) the maximum mass of rotating strange stars differs from
the static one by less than 4\% (the same result holds also for the e.o.s.
of strange matter with massive, interacting quarks
[Gondek-Rosi\'nska, et.~al 2000]).
}
\end{figure}

For interacting (but massless) quarks, through lowest order in gluon exchange,
the equation of state is identical to
that of non-interacting quarks \citep{chaplin},
$ p = (\rho - \rho_0) c^2 /3$,
 the only difference being in that the lower bound on
the density at zero pressure, following from conditions of neutron
stability (eq. [\ref{bmin}]), is decreased with respect
to the value for an ideal Fermi gas in a bag, by the same factor
as the bag constant:
$ \rho_0(\alpha_c) =
\left( 1 - {2\alpha_c/ \pi}\right)\rho_0(0)$.
Since the stellar mass scales as
$\rho_0^{-1/2}$, this implies that the least upper bound on the
mass of the star as a function of the QCD coupling constant is
given for non-rotating strange stars by  
$$  M_{max}(\alpha_c) =
\left( 1-{2\alpha_c\over \pi}\right)^{-1/2} M_{max}(0)\,, \eqno(3) 
$$
through first order in $\alpha_c$ \citep{Prakash}.
 For $\alpha_c =0.4$, eqs.~(2), (3) give a
maximum strange star mass of $ 2.29M_\odot$,
higher  by 16\% than the maximum mass which is obtained for $\alpha=0$.

Stellar rotation at a frequency up to 700 Hz would increase
the maximum mass by only a few percent. In Fig.~2, we present the maximum
mass of a strange star as a function of the rotational frequency,
$f=1/P$. This is not the frequency dependence of the mass of a single
star---what is plotted is the termination point, at each frequency,
of a sequence of stellar models varying in mass. Thus, the baryon number 
of the maximum-mass model varies with frequency. 
In fact, for all $f>0$, the maximum-mass stars are supramassive---if
spun down at constant baryon number to $f=0$, they would become unstable
to collapse (compare the discussion of neutron-star models in
Cook et al., 1994\nocite{1994ApJ...424..823C}).
We obtained the results
presented in Fig.~2 with an accurate code based on spectral methods,
developed by, and described in, Gourgoulhon et. al. (1999).

\begin{figure}
\includegraphics[width=\columnwidth]{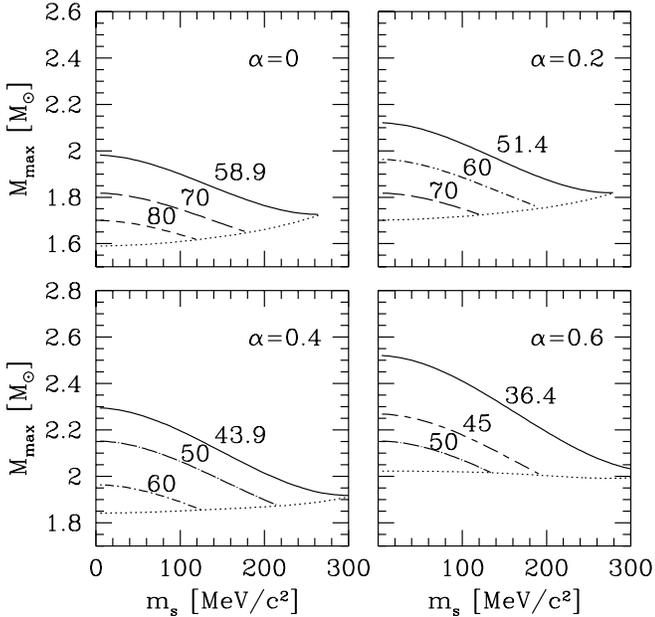}
\caption{The maximum mass of static strange stars in the MIT bag model,
 as a function of the strange quark mass, for various
values of the QCD coupling constant. $M_{max}$
is always above the  dotted line, which corresponds to the
upper bound on $B$ (Section 5); and below
the continuous line, which corresponds to the lower
bound on the bag constant,
 $B_{min}=58.9 (1-2\alpha_c/\pi)\,$MeV$\,{\rm fm}^{-3}$
(eq. [\ref{bmin}]).
Also shown are lines corresponding to other fixed values of the bag constant:
$B= 80,\,70,\,60,\,50\,$, and $45\,$MeV$\,{\rm fm}^{-3}$.
}
\end{figure}

\section{The maximum mass of non-rotating strange stars as a function of $m_s$
and $\alpha_c$}

To determine the maximum
mass of such stars for specific values of model parameters 
($B, \alpha_c, m_s$), the e.o.s. of matter composed of interacting,
massive quarks must be considered, and it
can only be determined
numerically. The maximum stellar mass then following
from the lower bound on $B$ implicit in eq. (\ref{bmin}) is
exhibited
in Fig. 3 (continuous lines) as a function of the strange-quark mass
and of the QCD coupling constant.

From Fig.~3 we can conclude for example, that if a strange star
of two solar masses ($2M_\odot$) were identified, and if
$\alpha_c$ were less than 0.4, then $m_s < 200\,$MeV
(with both quantities renormalized at $313\,$MeV).

If the density at zero pressure of quark matter is indeed close to
its lowest bound given by the stability limit of eq. (\ref{bmin}), then for
$\alpha_c > 0$ it can attain a value lower than $4\times
10^{14}$g\,cm$^{-3}$ and the mass of even a non-rotating strange star may be
sufficiently high to allow an orbital frequency of $1.06\,$kHz in
the marginally stable orbit. It would appear that in the model of
interacting quarks considered here, and in contrast to models in
which $\rho_0 > 4.2 \times 10^{14}$g\,cm$^{-3}$, the possible
presence of moderately rotating
strange stars in LMXBs could be compatible with the
keplerian model of kHz~QPOs---compare \citet{1999A&A...344L..71B}.

\section{The maximum value of $B$}

There is another stability constraint limiting the parameters of
quark matter, if such a substance
 is indeed the ground state of matter. The chemical
potential at zero pressure of quark matter in three flavours 
and electrons in beta equilibrium, should be less than the rest
energy per nucleon of the iron nucleus 
$\mu_0^{u,d,s} < \mu(^{56}{\rm Fe}) = 930.4 \,\mathrm{MeV}$
(Farhi and Jaffe 1984). For massless quarks this corresponds to 
$B/(1-2\alpha_c/\pi) < 91.49\,{\rm MeV}$,
 but in general the constraint depends on the strange quark mass,
and the maximum stellar mass corresponding to this upper bound on $B$
is exhibited in each panel of Fig.~3 as a dotted line
[for details see \cite{Prakash} and Zdunik et al. (2000)].
Note that the upper and lower constraints on $B$, when taken together,
exclude high masses of
the strange quark for low values of $\alpha_c$,
if strange quark gas is to be the stable form of matter. 

\section{Summary}

We have investigated the question whether moderately rotating strange
stars with  masses somewhat higher than
$2M_\odot$ are allowed by relativistic equations of stellar equilibrium,
and found that the answer could, in principle, be positive within the MIT
bag model of beta-stable quark matter.
An extension of this discussion to other models of strange matter
will be given elsewhere.

The physical constraints
on the bag constant following from the basic hypothesis of stability
of self-bound quark matter (eq. [\ref{bmin}]) allow $B$
to be so small, that the corresponding mass of the star could
be as high as $2.5M_\odot$ (eq. [3], Fig.~3).
However, the strange star mass cannot be higher than $2.6M_\odot$,
even for stars of rotational period as short as 1.6 ms.

The perturbative approach used here is sensible only if
the value of the QCD coupling constant $\alpha_c$ is small---we used 
$0\le\alpha_c\le0.6$. In this range, the direct dependence of $M_{max}$ on
the coupling is practically negligible (Fig.~1), but the window of allowed
values of $B$ does depend on $\alpha_c$ (Fig.~3). 
The actual value of $B$ is subject to a very large uncertainty.
Fits to the hadronic mass spectrum (DeGrand et al., 1975) gave
$B=59~{\rm MeV~fm^{-3}}$. 

The stellar mass
decreases with increasing mass, $m_s$, of the strange quark,
and is lower by $10\%$ to $20\%$
than the one for massless-quark matter
for the typical
range considered in the literature,
$150\le m_s c^2/{\rm MeV} \le 300$ (Madsen 1999).
 Finally, the maximum stellar mass
for strange stars rotating with a frequency up to  $700\,$Hz, is larger
than the one for non-rotating stars by less than 4\% (Fig.~2).

The physical reason for which large values of mass are obtained
for a low value of the bag constant, is that the latter is
proportional to the density of quark matter at zero pressure, $\rho_0$,
while
the maximum mass of the star is proportional to  $\rho_0^{-1/2}$.
Thus, within the MIT bag model of quark matter,
strange stars with mass $M>2.0 M_\odot$ must have
surface densities $\rho_0< \rho_1\equiv 4.2\times10^{14}\,$g/cm$^3$
(unless rotating with periods $\le1.5\,$ms).

\bigskip
This research was supported in part by KBN grants
2P03D00418, 2P03D02117, 2P03D04013.
The numerical calculations have been performed in part on
DARC computers purchased thanks to a special grant from the SPM
and SDU departments of CNRS. 

\newcommand{\apjl}{ApJ Let.}
\newcommand{\apj}{ApJ}
\newcommand{\mnras}{MNRAS}
\newcommand{\nat}{Nature}
\newcommand{\apjs}{ApJ Supp.}
\newcommand{\aap}{A\&A}
\newcommand{\iaucirc}{IAU Circ.}


\begin{thebibliography}{22}
\expandafter\ifx\csname natexlab\endcsname\relax\def\natexlab#1{#1}\fi

\bibitem[{Alcock} et~al.(1986){Alcock}, {Farhi}, and
  {Olinto}]{1986ApJ...310..261A}
{Alcock}, C., {Farhi}, E., and {Olinto}, A., 1986, \apj, 310, 261

\bibitem[{Alpar}(1987)]{Alpar}
Alpar, A. 1987, Phys. Rev. Lett. 58, 2152

\bibitem[{Arnett} and {Bowers}(1977)]{1977ApJS...33..415A}
{Arnett}, W.~D. and {Bowers}, R.~L., 1977, \apjs, 33, 415

\bibitem[{Bethe}, {Brown} and {Cooperstein}(1987)]{Bethe}
Bethe, H.~A., Brown, G.~E., and Cooperstein, J., 1987, Nucl. Phys. A462, 791

\bibitem[Bodmer(1971)]{1971Bodmer}
Bodmer, A.~R., 1971, Phys. Rev., 4, 1601

\bibitem[{Bonazzola} et~al.(1996){Bonazzola}, {Frieben}, and
  {Gourgoulhon}]{1996ApJ...460..379B}
{Bonazzola}, S., {Frieben}, J., and {Gourgoulhon}, E., 1996, \apj, 460, 379

\bibitem[{Brecher} and {Caporaso}(1976)]{1976Natur.259..377B}
{Brecher}, K. and {Caporaso}, G., 1976, \nat, 259, 377

\bibitem[{Bulik} et~al.(1999a){Bulik}, {Gondek-Rosi\'nska}, and
  {Klu\'zniak}]{1999A&A...344L..71B}
{Bulik}, T., {Gondek-Rosi\'nska}, D., and {Klu\'zniak}, W., 1999a,
 \aap, 344, L71

\bibitem[{Bulik} et~al.(1999b){Bulik}, {Gondek-Rosi\'nska}, and
  {Klu\'zniak}]{1999b}
{Bulik}, T., {Gondek-Rosi\'nska}, D., and {Klu\'zniak}, W., 1999b,
Astro. Lett. and Communications, 38, 77

\bibitem[{Caldwell} and {Friedman}(1991)]{Caldwell}
Caldwell, R.R., and Friedman, J.R., 1991, Physics Lett. B 264, 143

\bibitem[{Chapline} and {Nauenberg}(1976)]{chaplin}
{Chapline}, G. and {Nauenberg}, M. 1976, Nature, 264, 235

\bibitem[{Cheng} and {Dai}(1996)]{1996ApJ...468..819C}
{Cheng}, K.~S. and {Dai}, Z.~G., 1996, Phys. Rev. Lett., 80, 18

\bibitem[{Cook} et~al.(1994){Cook}, {Shapiro}, and
  {Teukolsky}]{1994ApJ...424..823C}
{Cook}, G.~B., {Shapiro}, S.~L., and {Teukolsky}, S.~A., 1994, \apj, 424, 823

\bibitem[Farhi (1991)]{Farhi1991}
{Farhi}, E., 1991, Nuc. Phys. B (Proc. Suppl.) 24B, 3

\bibitem[{Farhi} and {Jaffe}(1984)]{FarJaf}
{Farhi}, E., and Jaffe, R.L., 1984, Phys. Rev. D, 30, 2379

\bibitem[{Friedman} et~al.(1986){Friedman}, {Parker}, and
  {Ipser}]{1986ApJ...304..115F}
{Friedman}, J.~L., {Parker}, L., and {Ipser}, J.~R., 1986, \apj, 304, 115

\bibitem[{Frieman} and {Olinto}(1989)]{1989Natur.341..633F}
{Frieman}, J.~A. and {Olinto}, A.~V., 1989, \nat, 341, 633

\bibitem[{Glendenning}(1989)]{1989PhRvL..63.2629G}
{Glendenning}, N.~K., 1989, Physical Review Letters, 63, 2629

\bibitem[{Gondek-Rosi{\'n}ska} et.~al. (2000)]{Gon2000}
Gondek-Rosi{\'n}ska, D., et.~al. 2000, in preparation

\bibitem[{Gourgoulhon} et~al.(1999){Gourgoulhon}, {Haensel}, {Livine},
  {Paluch}, {Bonazzola}, and {Marck}]{1999A&A...349..851G}
{Gourgoulhon}, E., {Haensel}, P., {Livine}, R., {Paluch}, E., {Bonazzola}, S.,
  and {Marck}, J.~A., 1999, \aap, 349, 851

\bibitem[Haensel(1987)]{Haen1987}
{Haensel}, P., 1987, Acta Phys. Polon. B18, 739

\bibitem[{Haensel} and {Zdunik}(1990)]{1990A&A...229..117H}
{Haensel}, P. and {Zdunik}, J.~L., 1990, \aap, 229, 117

\bibitem[{Haensel} et~al.(1986){Haensel}, {Zdunik}, and
  {Schaeffer}]{1986A&A...160..121H}
{Haensel}, P., {Zdunik}, J.~L., and {Schaeffer}, R., 1986, \aap, 160, 121

\bibitem[{Klu{\'z}niak}(1994)]{Kluz94}
Klu\'zniak, W. 1994, A\&A 286, L17

\bibitem[{Klu{\'z}niak}(1998)]{1998ApJ...509L..37K}
{Klu{\'z}niak}, W., 1998, \apjl, 509, L37

\bibitem[{Klu\'zniak} et~al.(1990){Klu\'zniak}, {Michelson}, and
  {Wagoner}]{1990ApJ...358..538K}
{Klu\'zniak}, W., {Michelson}, P., and {Wagoner}, R.~V., 1990, \apj, 358, 538

\bibitem[{Madsen}(1988)]{Madsen}
{Madsen}, J., 1988, Phys. Rev. Lett. 61, 2909

\bibitem[{Madsen}(1999)]{mad99}
{Madsen}, J., in Hadrons in dense mater and hadrosynthesis,
 Cleymans, J., ed., Lecture Notes in Physics, (Springer), in press,
 astro-ph/9809032

\bibitem[{Miralda-Escud\'e} et~al.(1990){Miralda-Escude}, {Paczy\'nski}, and
  {Haensel}]{1990ApJ...362..572M}
{Miralda-Escud\'e}, J., {Paczy\'nski}, B., and {Haensel}, P., 1990, 
\apj, 362, 572

\bibitem[{Oppenheimer} and {Volkoff}(1939)]{1938PhysRev...55..374}
{Oppenheimer}, J.~R. and {Volkoff}, G.~M., 1939, Phys. Rev., 55, 374

\bibitem[{Prakash} et~al.(1990){Prakash}, {Baron}, and {Prakash}]{Prakash}
{Prakash}, M., {Baron}, E., and {Prakash}, M., 1990, Phys. Letters., 243, 175

\bibitem[{Salgado}, et.~al.(1994)]{Salgado}
{Salgado}, M., {Bonazzola}, S., {Gourgoulhon}, E., and {Haensel}, P., 1994,
\aap, 291, 155

\bibitem[{Stergioulas} et~al.(1999){Stergioulas}, {Klu{\'z}niak}, and
  {Bulik}]{1999A&A...352L.116S}
{Stergioulas}, N., {Klu{\'z}niak}, W., and {Bulik}, T., 1999, \aap, 352, L116

\bibitem[{Strohmayer} et~al.(1996){Strohmayer}, {Zhang}, and
  {Swank}]{1996IAUC.6320....1S}
{Strohmayer}, T., {Zhang}, W., and {Swank}, J., 1996, \iaucirc, 6320, 1

\bibitem[{Strohmayer} et~al.(1997){Strohmayer}, {Jahoda}, {Giles}, and
  {Lee}]{1997ApJ...486..355S}
{Strohmayer}, T.~E., {Jahoda}, K., {Giles}, A.~B., and {Lee}, U., 1997, \apj,
  486, 355

\bibitem[{van der Klis}(1998)]{QPOrev98}
{van der Klis}, M., 1998, in Proceedings of the Third William Fairbank Meeting,
  Rome, astro-ph/9812395

\bibitem[{van der Klis} et~al.(1996){van der Klis}, {Swank}, {Zhang}, {Jahoda},
  {Morgan}, {Lewin}, {Vaughan}, and {Van Paradijs}]{1996IAUC.6319....1V}
{van der Klis}, M., {Swank}, J., {Zhang}, W., {Jahoda}, K., {Morgan}, E.,
  {Lewin}, W., {Vaughan}, B., and {van Paradijs}, J., 1996, \iaucirc, 6319, 1

\bibitem[{Wijnands} and {van der Klis}(1998)]{1998Natur.394..344W}
{Wijnands}, R. and {van der Klis}, M., 1998, \nat, 394, 344

\bibitem[Witten(1984)]{1984Witten}
Witten, E., 1984, Phys. Rev., 30, 272

\bibitem[{Zdunik} (2000)]{Zdun2000}
Zdunik, J.L, et~al., 2000, in preparation

\end{thebibliography}
\end{document}